%% file: main.tex
\documentclass[5p,longtitle,sort&compress,reversenotenum]{elsarticle}

\usepackage{amssymb}
\usepackage{amsmath}
\usepackage{booktabs}
\usepackage{multirow}

\usepackage[range-phrase=-, range-units=single]{siunitx}
\journal{Physics Letters B}

\begin{document}

\begin{frontmatter}

\title{Hint for a TeV neutrino emission from the Galactic Ridge with ANTARES}
\input{authors.tex}

\begin{abstract}
Interactions of cosmic ray protons, atomic nuclei, and electrons in the interstellar medium in the inner part of the Milky Way produce a $\gamma$-ray flux from the Galactic Ridge. If the $\gamma$-ray emission is dominated by proton and nuclei interactions, a neutrino flux comparable to the $\gamma$-ray flux is expected from the same sky region. \\
Data collected by the ANTARES neutrino telescope are used to constrain the neutrino flux from the Galactic Ridge in the \SIrange{1}{100}{\tera\electronvolt} energy range. Neutrino events reconstructed both as tracks and showers are considered in the analysis and the selection is optimized for the search of an excess in the region $|l|<\SI{30}{\degree}$, $|b|<\SI{2}{\degree}$. The expected background in the search region is estimated using an off-zone region with similar sky coverage. Neutrino signal originating from a power-law spectrum with spectral index ranging from $\Gamma_\nu=1$ to $4$ is simulated in both channels. The observed energy distributions are fitted to constrain the neutrino emission from the Ridge. \\
The energy distributions in the signal region are inconsistent with the background expectation at $\sim 96\%$ confidence level. The mild excess over the background is consistent with a neutrino flux with a power law with a spectral index $2.45^{+0.22}_{-0.34}$ and a flux normalization $\frac{dN_\nu}{dE_\nu} = 4.0^{+2.7}_{-2.0} \times \SI{e-16}{\per\giga\electronvolt\per\square\centi\meter\per\second\per\steradian}$ at \SI{40}{\tera\electronvolt} reference energy. Such flux is consistent with the expected neutrino signal if the bulk of the observed $\gamma$-ray flux from the Galactic Ridge originates from interactions of cosmic ray protons and nuclei with a power-law spectrum extending well into the PeV energy range.
\end{abstract}


\end{frontmatter}

\section{Introduction}
\label{sec:intro}

The cosmic ray content of the Milky Way is determined by the star evolution process that proceeds at different rates in different parts of the Galactic Disk. The cosmic ray spectrum measured locally is approximately a power law $dN/dE\propto E^{-\Gamma}$ with an average spectral index $\Gamma\simeq 2.7$ below the knee, a feature in the PeV range, and an average spectral index $\Gamma \simeq 3.0$ above this feature \cite{KASCADEGrande:2011kpw,IceCube:2013ftu}. Up to \SI{e18}{\electronvolt}, cosmic rays seem mainly of Galactic origin \cite{Grenier:2015egx}. The cosmic ray spectrum may depend on a subtle balance between the rate of injection of ``fresh'' cosmic rays from currently unknown sources and the rate of escape of ``old'' cosmic rays diffusing through the Galactic magnetic field \cite{Berezinskii:1990,Blasi:2013rva} of yet uncertain geometry \cite{Jansson:2012pc,Pshirkov:2011um}. According to this description, the locally measured spectrum is not necessarily representative of that present in the whole Galaxy. Its power-law spectral index, presumably regulated by the average spectral index of the injection spectrum from the sources and the energy dependence of the diffusion coefficient, can vary depending on the source population properties and on the structure of the magnetic field in the interstellar medium that may lead to variations of the diffusion coefficient with the distance from the Galactic Center \cite{Erlykin:2012dp,2015ApJ...815L..25G}. The origin of the cosmic ray knee feature is uncertain: it can be related to the average maximal energy attainable by the particle accelerators operating in Galactic sources \cite{Thoudam:2016syr}, or related to the change of cosmic ray propagation regime through the interstellar medium \cite{Giacinti:2015hva}. Finally, the knee could originate from a local cosmic ray spectrum feature imprinted by a single nearby source \cite{Erlykin:1997bs}.

The cosmic ray spectrum from different parts of the Galactic Disk can be constrained using $\gamma$-ray and neutrino observations. Interactions of cosmic rays in the interstellar medium lead to the $\gamma$-ray glow of the disk of the Milky Way \cite{Strong:2007nh,Ackermann:2012}. The diffuse $\gamma$-ray emission from the Galaxy is detected by Fermi Large Area Telescope (LAT) \cite{Atwood:2019} up to several \si{\tera\electronvolt} energy \cite{Neronov:2019ncc}. Analysis of the $\gamma$-ray data indicates that the spectrum of cosmic rays in the inner Galaxy may be harder than the locally measured spectrum. Its spectral index may be as hard as $\Gamma \simeq 2.4$ in the innermost part of the disk, the Galactic Ridge \cite{Neronov:2013lza,Neronov:2015vua,Yang:2016jda,Fermi-LAT:2016zaq,2017PhRvL.119c1101G}. In galactic coordinates, this region extends over galactic longitudes $l$ with $|l|<l_{\rm ridge}$, and galactic latitudes $b$ with $|b|<b_{\rm ridge}$, where $l_{\rm ridge} \sim \SIrange{30}{40}{\degree}$, $b_{\rm ridge} \sim \SIrange{2}{3}{\degree}$. This paper focuses on the region with $l_{\rm ridge} = \SI{30}{\degree}$ and $b_{\rm ridge} = \SI{2}{\degree}$ to allow direct comparisons with the Fermi-LAT $\gamma$-ray measurements reported in \cite{Neronov:2019ncc}. This also allows concentrating the efforts on a smaller region where the eventual signal is less likely to be diluted in the expected background.

Cosmic ray interactions in the Galactic Ridge are also expected to generate a neutrino flux, with spatial morphology and spectrum similar to that of the $\gamma$-ray signal \cite{Neronov:2013lza,Kachelriess:2014oma,Neronov:2014uma}. A search for neutrino emission from the Galactic Ridge direction was previously reported by ANTARES \cite{Adrian-Martinez:2016fei}, considering the region $(|l|<\SI{40}{\degree}, |b|<\SI{3}{\degree})$. This prior search was performed using events induced by charged-current muon neutrino interactions (track events), and limited to reconstructed energies above \SI{10}{\tera\electronvolt}. The upper limit on the neutrino flux was estimated to be at the level of $\num{6.0e-5} (E_\nu/\SI{1}{\giga\electronvolt})^{-\Gamma_\nu} \, \si{\per\giga\electronvolt\per\square\centi\meter\per\second\per\steradian}$ for the assumed spectral index of the power-law neutrino spectrum $\Gamma_\nu=2.5$ and somewhat different normalizations for other spectral indices. The ANTARES limit has imposed a tight constraint on the neutrino flux at the level close to the extrapolation of the Fermi/LAT spectrum of the Galactic Ridge to the energy range around \SI{100}{\tera\electronvolt}. Other recent searches are mostly extended to the full Galactic Plane region e.g., with ANTARES data \cite{ANTARES:2017nlh}, IceCube data \cite{IceCube:2017trr}, and the combination of both \cite{ANTARES:2018nyb}.

The analysis of \cite{Adrian-Martinez:2016fei} used ANTARES data collected before December 2013 and information on the instrument calibrations available at the moment of publication. The work presented in this paper reports an update on the search for the neutrino signal from the Galactic Ridge taking into account several changes. A larger data sample has been used and, in addition to muon neutrinos reconstructed as tracks, showering events induced by electron neutrinos and neutral current interactions are also included. Finally, a better understanding of the instrument with refined calibrations, and updated energy estimators for the track events, allows an improved quality of the analyzed data sample.

\section{Data analysis}

The updated Galactic Ridge analysis uses ANTARES data collected between May 2007 and February 2020 for tracks and up to December 2020 for showering events. The data set exposure is 1.6 times larger than that used in the latest analysis of the ANTARES collaboration \cite{Adrian-Martinez:2016fei}.

As described in section \ref{sec:intro}, the Galactic Ridge is defined as the region limited by longitude and latitude ranges $|l|<\SI{30}{\degree}$ and $|b|<\SI{2}{\degree}$ \cite{Neronov:2019ncc}. The data are separated into two samples: track events which correspond to muons produced in $\nu_\mu$ charged-current interactions and shower events which are mainly associated with $\nu_e$ charged-current interactions and neutral-current interactions. Tau neutrino interactions also contribute to both channels.

For the track sample, only events with reconstructed direction $|l|<\SI{30}{\degree}$, $|b|<\SI{2}{\degree}$ are selected and the same quality cuts as in \cite{Adrian-Martinez:2016fei} are imposed. This corresponds to selecting only events with values of the parameter estimating the precision of the reconstructed direction better than $\beta_{\rm cut}=\SI{0.5}{\degree}$, and the parameter estimating the quality of track reconstruction above $\Lambda_{\rm cut}=-5.0$.

The selection of shower-like events is similar to the one presented by \cite{ANTARES:2021cwc}. As the shower angular resolution is not as good as for tracks, the search region is extended to $|l|<\SI{33}{\degree}$, $|b|<\SI{5}{\degree}$ to maximize the sensitivity to the neutrino signal in the Galactic Ridge (this extension has been optimized to maximize the signal acceptance for an $E^{-2.4}$ spectrum).

\begin{figure}[hbtp]
    \centering
    \includegraphics[width=\linewidth]{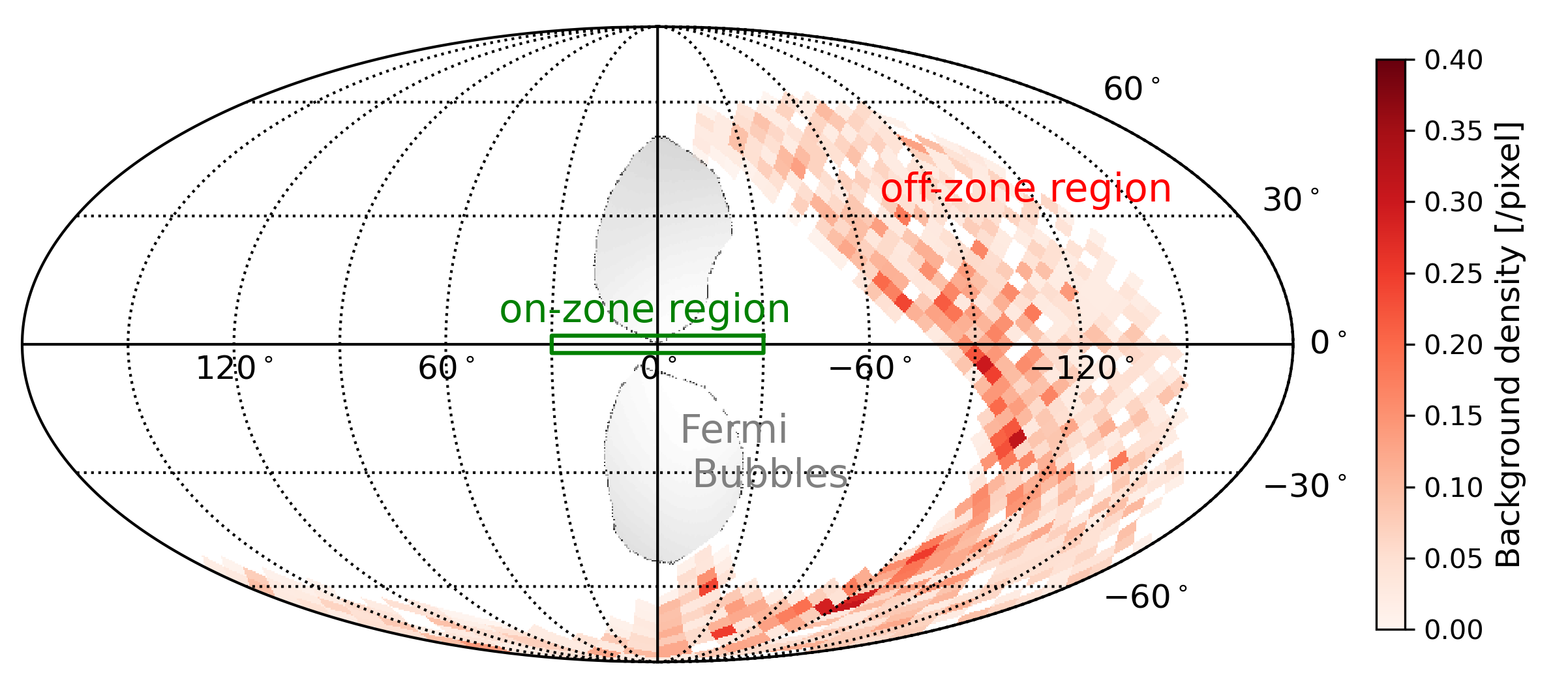}
    \caption{Illustration of the on-off analysis technique for the track channel in the galactic coordinates. The green box delimits the on-zone region $(|l|<\SI{30}{\degree}, |b|<\SI{2}{\degree})$, while the red pixels correspond to the density of events from the off-zone region used for the background estimate. The gray areas represent the Fermi bubbles.}
    \label{fig:offregion}
\end{figure}

For both track and shower samples, the background in the signal region is estimated by using the off-zone region with the same sky coverage but shifted in right ascension (RA). Each neutrino event in real data that does not overlap with either the Galactic Ridge or the Fermi bubbles\footnote{The Fermi bubbles are cautiously excluded from the off-zone estimation as they may host a neutrino flux \cite{Lunardini:2011br} which is not investigated in this search.} is shifted randomly several times in RA. If an event $i$ enters into the signal region, it is used for the background estimate, with weight $w_i$ proportional to how often it has fallen into the signal region out of all the trials. All the events with $w_i > 0$ are then used to estimate the background: $B = \sum w_i$. The statistical error on this estimate is computed as $\sigma_B =\sqrt{\sum w_i^2}$. An illustration of the off-zone region for the track analysis is shown in Fig.~\ref{fig:offregion}.

The event selection considered in the analysis uses a revised energy estimate compared to \cite{Adrian-Martinez:2016fei} for the track sample (with a time-dependent correction to account for performance changes over time) and the same estimator as in \cite{ANTARES:2021cwc} for showers. The former is only a rough estimate of the original neutrino energy and may be off by up to one order of magnitude for individual events, while the latter has an intrinsic resolution of $5-10\%$ \cite{ANTARES:2017ivh}.

The neutrino signal is simulated using the standard run-by-run ANTARES Monte Carlo (MC) simulations \cite{ANTARES:2020bhr} and assuming a simple power-law neutrino spectrum:

\begin{equation}
    \Phi(E) = \frac{dN_\nu}{dE_\nu} = \Phi_0 \left(\frac{E_\nu}{E_0}\right)^{-\Gamma_\nu},
\end{equation}
where $\Gamma_\nu$ is the spectral index of the power law and $\Phi_0 = \Phi(E_0)$ is the normalization of the neutrino flux for a single flavour (the total neutrino flux is $3\times \Phi_0$, assuming total mixing of neutrino flavours due to neutrino oscillations during the propagation of the signal from the source to the Earth). The energy normalization has been fixed for convenience to $E_0=\SI{40}{\tera\electronvolt}$.

The results are interpreted in terms of the reconstructed energy $E_{\rm rec}$ distribution for tracks and showers separately, by comparing the observation in the defined search region to the background estimated from the off-zone region. For each event category, six bins are defined uniformly in logarithmic scale between $\log_{10}(E_{\rm rec}) = 2$ and $5$. The following likelihood is defined:

\begin{equation}
    \small \mathcal{L}\left(\{N_i\}; \{S_i^{(\Gamma_\nu)}\}, \{B_i\}, \Phi_0 \right) = \prod_{i=1}^{12} \textrm{Poisson}\left(N_i, B_i + \Phi_0 S_i^{(\Gamma_\nu)}\right),
\end{equation}
where $N_i$ is the observed number of events in bin $i$, $B_i$ is the corresponding expected background, $S_i^{(\Gamma_\nu)}$ is the signal prediction for $\Phi_0 = \SI{1}{\per\giga\electronvolt\per\square\centi\meter\per\second\per\steradian}$ and a spectral index $\Gamma_\nu$, and the product runs over the twelve bins (6 for tracks, 6 for showers).

A Bayesian treatment is applied, where statistical and systematic uncertainties on the background and on the signal estimates are included as Gaussian priors $\pi(\{B_i\})$ and $\pi(\{S_i^{(\Gamma_\nu)}\})$. As the background is estimated from data, only the related statistical uncertainty is taken into account. For the signal, the MC statistical error is negligible and an overall 20\% normalization systematic uncertainty is included, as already prescribed in \cite{Adrian-Martinez:2016fei}. A flat prior $\pi(\Phi_0, \Gamma_\nu) \propto 1$ is assumed for the parameters of interest $\Phi_0$ and $\Gamma_\nu$ (with $1 \leq \Gamma_\nu \leq 4$).

The marginalized posterior distribution $P(\Phi_0, \Gamma_\nu)$ is obtained by factoring in the likelihood and the priors, and then integrating over the nuisance parameters:

\begin{align}
    P(\Phi_0, \Gamma_\nu) = &\int \mathcal{L}\left(\{N_i\}; \{S_i^{(\Gamma_\nu)}\}, \{B_i\}, \Phi_0 \right) \nonumber \\
    &\times \pi(\{B_i\}) \times \pi(\{S_i^{(\Gamma_\nu)}\}) \times \pi(\Phi_0, \Gamma_\nu) \nonumber \\
    &\times \prod_i \left( {\rm d}B_i {\rm d}S_i^{(\Gamma_\nu)} \right). 
\end{align}
Several outputs can be extracted, such as the best-fit values, 2D contours in the $(\Phi_0, \Gamma_\nu)$ plane, and best fit/ranges on $\Phi_0$ for a given spectral index.

The background distributions, as reported in Fig.~\ref{fig:spectra}, are used to generate background-only pseudo-experiments. The ANTARES sensitivity to the diffuse flux in the Galactic Ridge is then defined as the median upper limit coming from such pseudo-experiments. The related sensitivities for discrete values of the neutrino spectral index $\Gamma_\nu$ are presented in Tab.~\ref{tab:sens}. It shows that the inclusion of the shower sample allows improving the expected sensitivity by $20-30\%$ with respect to the case where only track-like events are considered.

\begin{table}[hbtp]
    \centering
    \caption{Sensitivity at 90\% C.L. on $\Phi(\SI{1}{\giga\electronvolt})$ for varying spectral indices, using only ANTARES track sample, only showers, or both.}
    \begin{tabular}{c|S[table-format=1.1e2]|S[table-format=1.1e2]|S[table-format=1.2e2]}
    \toprule
        \multirow{2}{*}{Spectral index} & \multicolumn{3}{c}{Sensitivity [\si{\per\giga\electronvolt\per\square\centi\meter\per\second\per\steradian}]}  \\
         & {Tracks} & {Showers} & {Combined} \\
    \midrule
        $\Gamma_\nu = 2.4$ & 5.0e-05 & 8.9e-05 & 4.0e-05 \\
        $\Gamma_\nu = 2.5$ & 1.4e-04 & 2.2e-04 & 1.1e-04 \\
        $\Gamma_\nu = 2.6$ & 3.7e-04 & 5.6e-04 & 2.8e-04 \\
        $\Gamma_\nu = 2.7$ & 9.5e-04 & 1.4e-03 & 7.3e-04 \\
        $\Gamma_\nu = 2.8$ & 2.4e-03 & 3.2e-03 & 1.8e-03 \\
    \bottomrule
    \end{tabular}
    \label{tab:sens}
\end{table}

\section{Results}

\begin{figure}[hbtp]
    \centering
    \includegraphics[width=\linewidth]{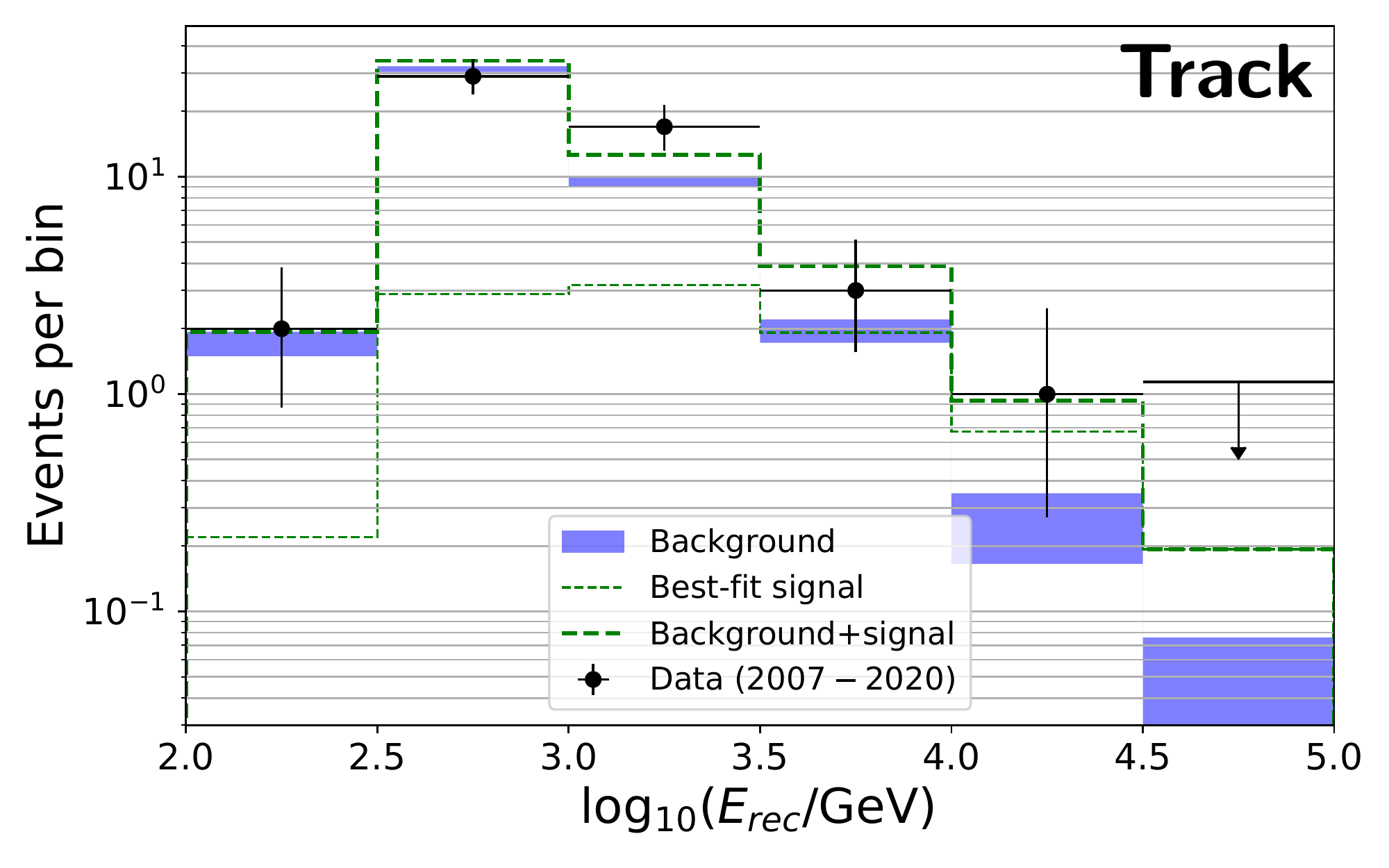}
    \includegraphics[width=\linewidth]{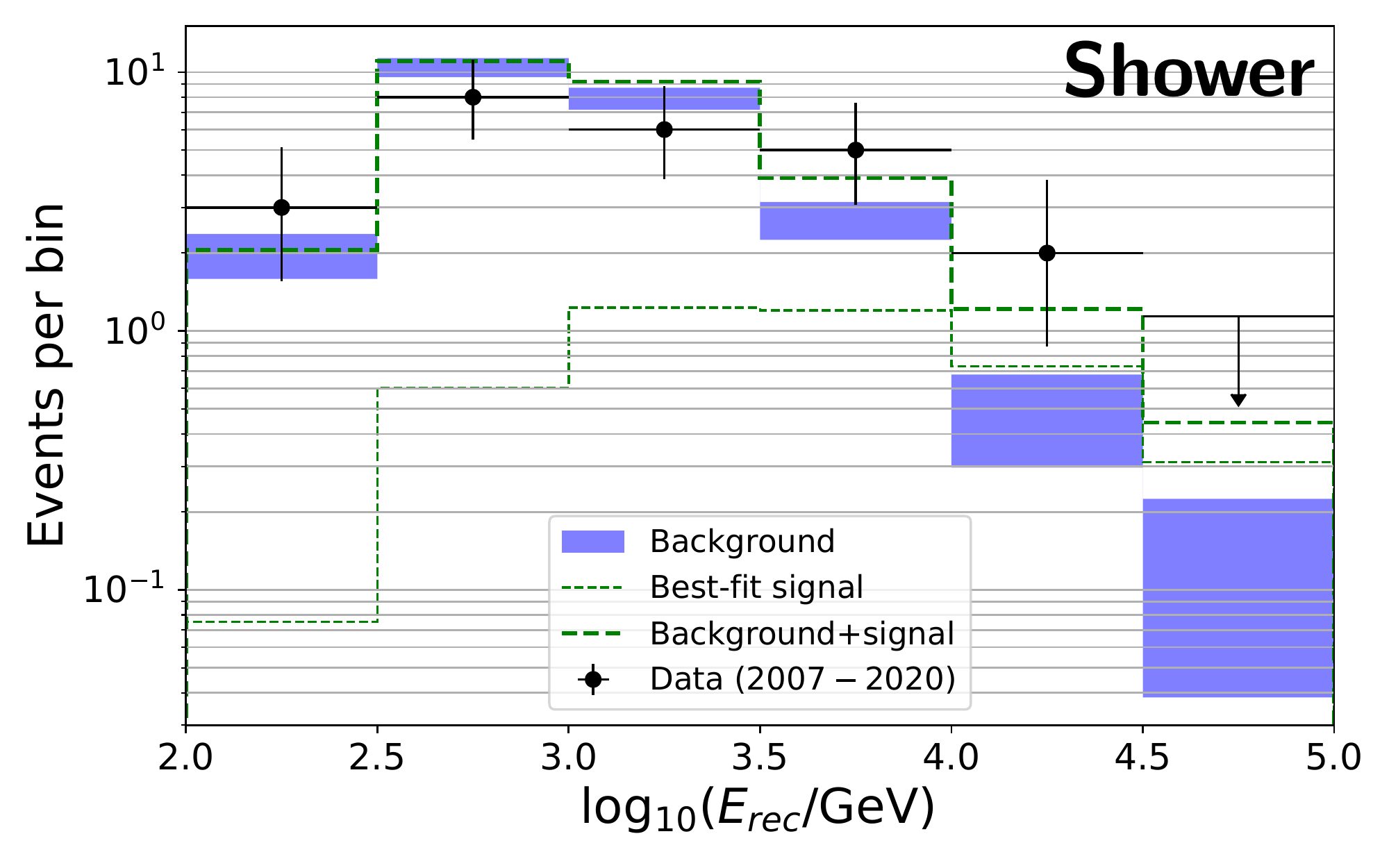}
    \caption{Reconstructed energy distribution for ANTARES track (top) and shower (bottom) samples. The black dots represent the observation in the search region with its statistical errors at 68\% C.L., with the ANTARES dataset spanning from 2007 to 2020. The blue histogram illustrates the expected background estimated using the off-zone region, the vertical bands representing the corresponding statistical uncertainty. The thin dashed green line shows the best-fit Galactic neutrino signal from Eq.~\ref{eq:bestfit} and the thicker dashed green line is the sum of this best-fit signal and the background.}
    \label{fig:spectra}
\end{figure}

Fig.~\ref{fig:spectra} shows the reconstructed energy distributions in the search region, the related background expectation, and the best-fit signal reported below. An excess is visible in the track and shower channels for high reconstructed event energies. Counting the number of events above \SI{1}{\tera\electronvolt}, 21 (13) events in the track (shower) channel, and comparing to the expected background, $11.7 \pm 0.6$ ($11.2 \pm 0.9$), provides a background rejection significance of 98\% (56\%), which corresponds to a $2.2\sigma$ ($0.2\sigma$) one-tailed excess.

\begin{figure}[hbtp]
    \centering
    \includegraphics[width=\linewidth]{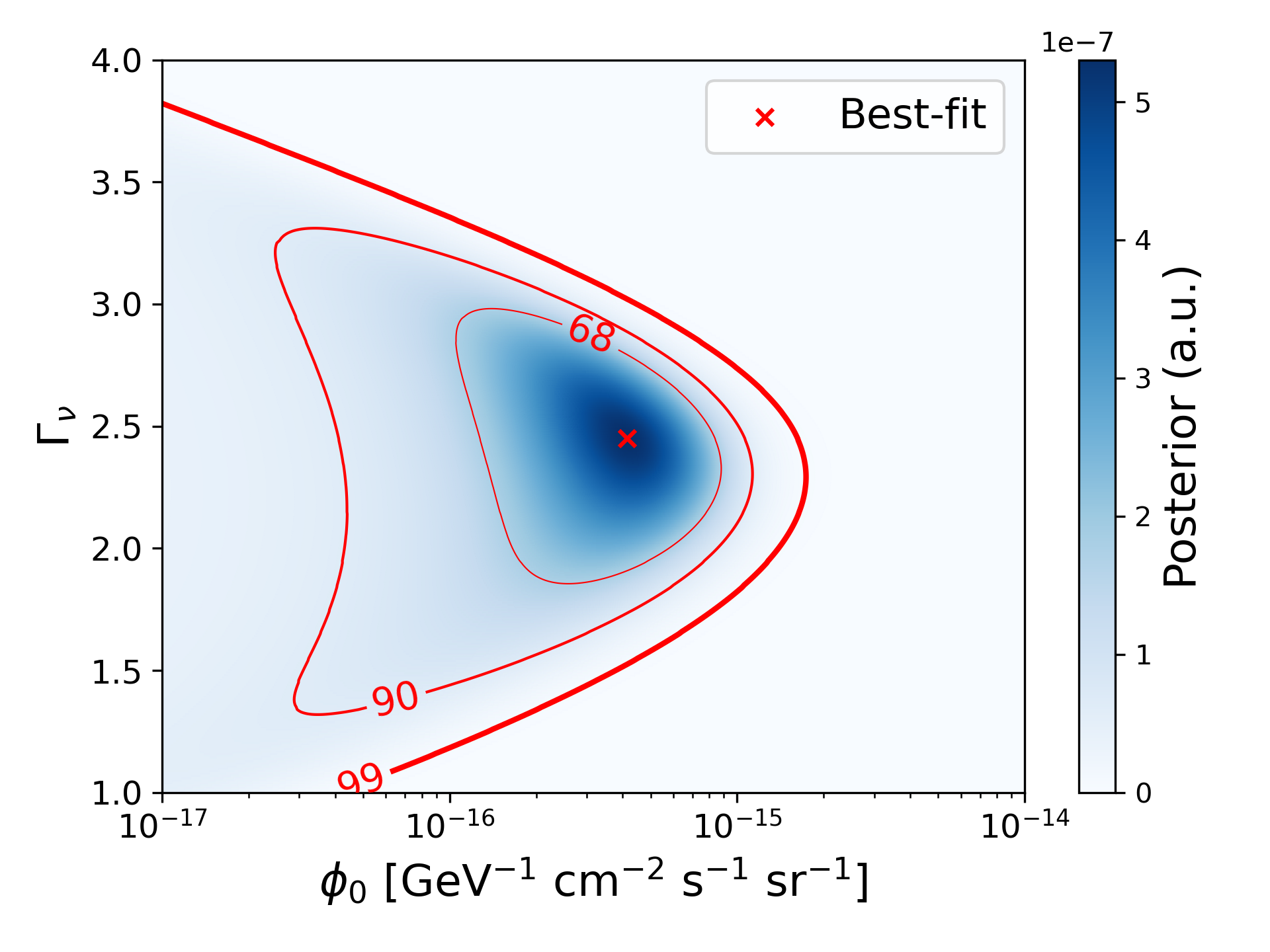}
    \caption{Marginalized posterior distribution in the plane $(\Phi(\SI{40}{\tera\electronvolt}) = \Phi_0, \Gamma_\nu)$. The red lines show the contours containing 68\%/90\%/99\% of the probability, and the best-fit point is indicated by the cross.}
    \label{fig:posterior}
\end{figure}

The Bayesian statistical analysis described in the previous section gives the 2D posterior distribution shown in Fig.~\ref{fig:posterior}. Several conclusions can be obtained from the latter:
\begin{itemize}
    \item The best fit of tracks+showers data corresponds to a per-flavour flux:
        \begin{align}
            \Phi(\SI{1}{\giga\electronvolt}) &= 7.6^{+5.0}_{-3.9} \times \SI{e-5}{\per\giga\electronvolt\per\square\centi\meter\per\second\per\steradian}, \nonumber \\
            \Phi(\SI{40}{\tera\electronvolt}) &= 4.0^{+2.7}_{-2.0} \times \SI{e-16}{\per\giga\electronvolt\per\square\centi\meter\per\second\per\steradian}, \nonumber\\
            \Gamma_\nu &= 2.45^{+0.22}_{-0.34}.
            \label{eq:bestfit}
        \end{align}

    \item Profiling to the best-fit spectral index $\Gamma_\nu = 2.45$, the 90\% credible interval ranges from $\Phi(\SI{1}{\giga\electronvolt}) = \num{1.6e-5}$ to $\SI{1.7e-4}{\per\giga\electronvolt\per\square\centi\meter\per\second\per\steradian}$.

    \item The background-only hypothesis is rejected at 96\% confidence level ($2.0\sigma$).
\end{itemize}

Further checks have been performed to ensure that these results remain stable against the methods used to estimate the background from the off-zone region and the energy of track events.

\begin{figure}[hbtp]
    \centering
    \includegraphics[width=\linewidth]{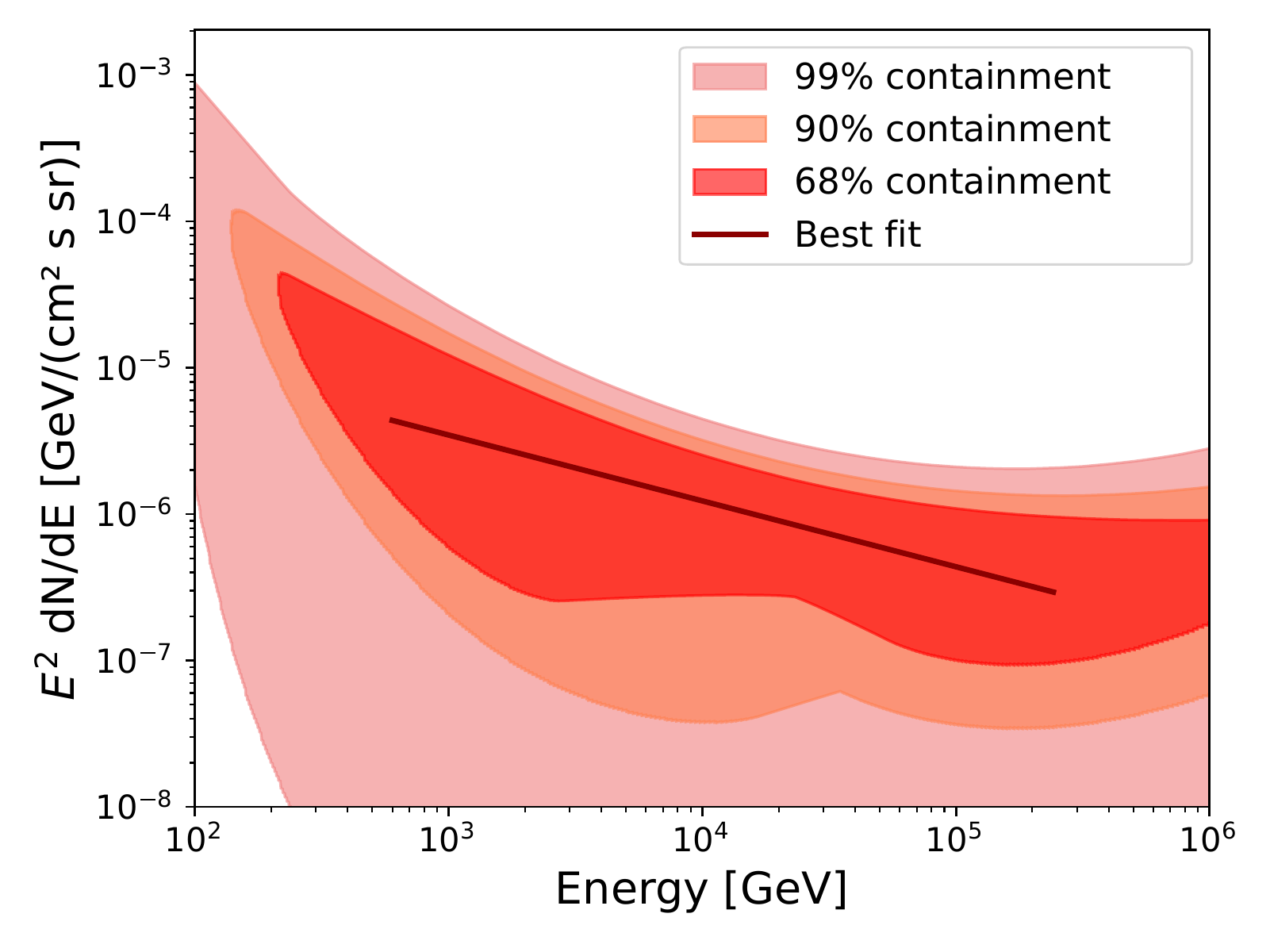}
    \caption{Constraints on the per-flavour neutrino energy-flux $E^2 \Phi(E)$ in the Galactic Ridge as a function of neutrino energy. The red-shaded bands show the envelopes of the 68\%/90\%/99\% constraints as shown in Fig.~\ref{fig:posterior} and the dark red line represents the best-fit flux. The endpoints on the x-axis illustrate the central energy ranges where 90\% of the considered neutrino signal is located for the various power-law spectra probed in the search.}
    \label{fig:constraints}
\end{figure}

Scanning over the points contained in the 68\%/90\%/99\% contours, Fig.~\ref{fig:constraints} shows the envelopes of the corresponding constraints in terms of energy-flux $E^2 \Phi(E)$.

The results are mainly driven by the observations of track-like events and thus by muon neutrino interactions. Additionally, as ANTARES track and shower samples only extend up to tens of \si{\tera\electronvolt}, the current data are not sufficient to conclude on the existence of an energy cutoff in the neutrino spectrum. In other words, the fit is insensitive to the value of such cutoff in the relevant range $\SI{100}{\tera\electronvolt} - \si{\peta\electronvolt}$. Additionally, given the size of the expected signal in the shower sample and that no significant excess is observed in this sample alone, the analysis cannot conclude on the relevance of the assumed equipartition between the three neutrino flavours.

To compare with previously reported ANTARES upper limits, the results with the larger region $(|l|<\SI{40}{\degree}, |b|<\SI{3}{\degree})$ have also been checked. The excess is no longer as important, as expected since the signal is diluted in a larger background, and corresponds to $1.3\sigma$ with a 90\% upper limit $\Phi(\SI{1}{\giga\electronvolt}) < \SI{1.2e-04}{\per\giga\electronvolt\per\square\centi\meter\per\second\per\steradian}$ for $\Gamma_\nu = 2.5$. The related sensitivity is $\SI[allow-number-unit-breaks=true]{6.7e-5}{\per\giga\electronvolt\per\square\centi\meter\per\second\per\steradian}$, which is worse than the one reported in \cite{Adrian-Martinez:2016fei} (\num{6.0e-5} in the same units), though a sensitivity improvement was expected given the increased statistics and the inclusion of shower-like events. The difference originates from the updated energy estimate that is now correctly accounting for the detector evolution over the years, and from issues in the signal Monte Carlo used in the previous analysis that was not properly restricted to the Galactic Ridge region, hence leading to a 20\% overestimate of the sensitivities in this past analysis.

\section{Discussion}
\label{sec:discussion}

The ANTARES observations reported in this article hint towards the existence of a neutrino flux from the Galactic Plane, and hence support the conventional interpretation of the previously observed $\gamma$-ray signal as being due to cosmic ray interactions with the interstellar medium. This updated analysis, with increased ANTARES exposure, in the direction of the Galactic Ridge reveals excesses in the energy distribution of both track and shower events that are inconsistent with the background-only hypothesis at 96\% confidence level. The analysis presented here is complementary to the searches in the Galactic Plane \cite{ANTARES:2017nlh,IceCube:2017trr,ANTARES:2018nyb}, though direct comparisons are not possible due to the different sizes of the probed regions.

\begin{figure}
    \centering
    \includegraphics[width=\linewidth]{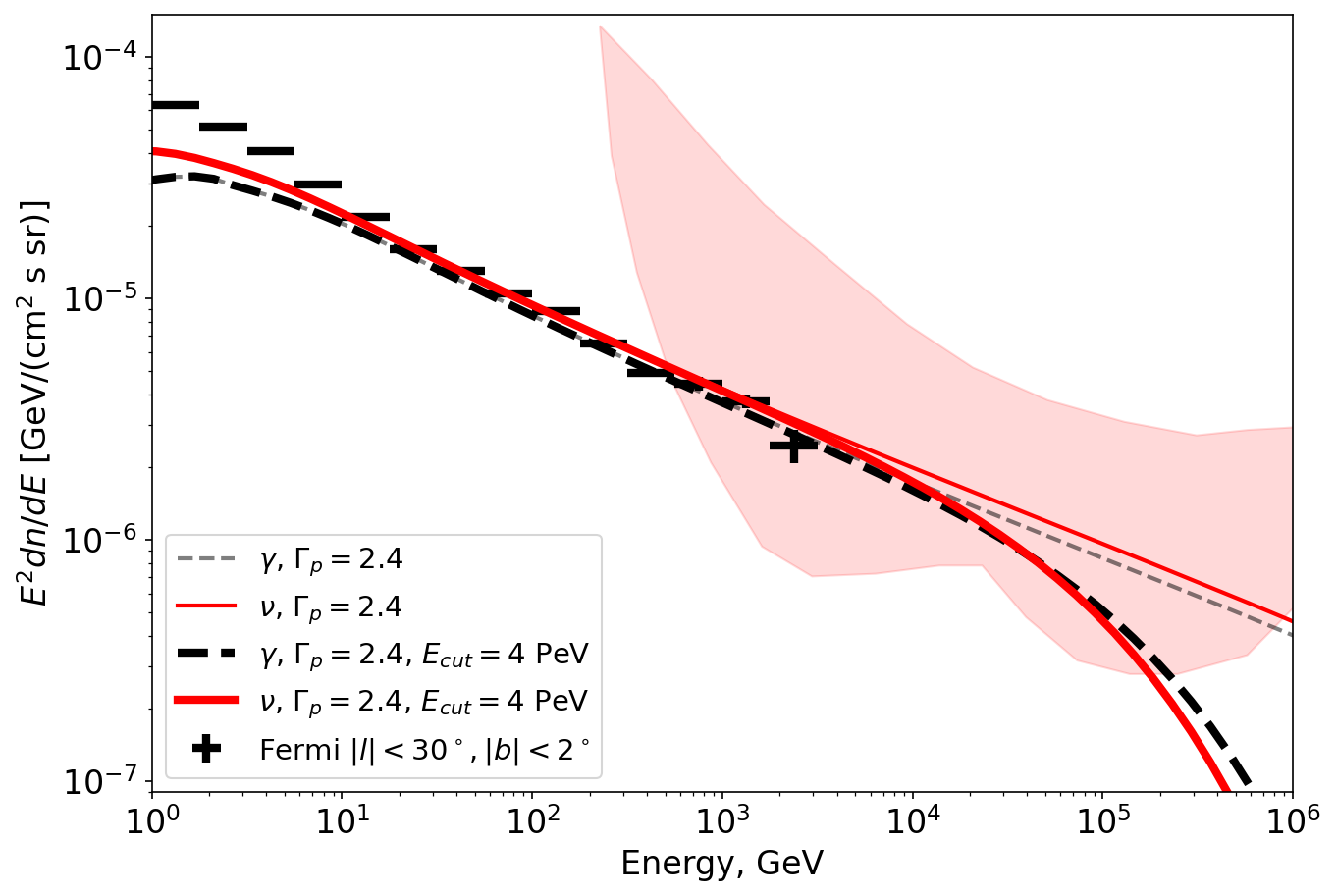}
    \caption{All-flavour neutrino flux corresponding to the 68\% containment contour for the ANTARES excess (red shading) compared to the Fermi/LAT diffuse  $\gamma$-ray flux (black data points) from the region $|l|<\SI{30}{\degree}$, $|b|<\SI{2}{\degree}$ \cite{Neronov:2019ncc}. Curves show model neutrino (solid) and $\gamma$-ray (dashed) pion decay spectra for different cosmic ray proton spectra: a power-law spectrum with a spectral index $\Gamma_p=2.4$ (thin) and one with the same spectral index and high-energy cut-off at $E_{cut}=\SI{4}{\peta\electronvolt}$ (thick).}
    \label{fig:gamma_nu}
\end{figure}

The excess of events in the Galactic Ridge direction is consistent with an estimate based on the $\gamma$-ray data. This consistency is illustrated in Fig.~\ref{fig:gamma_nu} where the $\gamma$-ray measurements of the diffuse flux from the region $|l|<\SI{30}{\degree}$, $|b|<\SI{2}{\degree}$ \cite{Neronov:2019ncc} are compared to the neutrino flux estimate derived above. The $\gamma$-ray spectrum alone is well fit by the model of $\pi^0$ decays from interactions of protons with a rather hard power-law spectrum with a spectral index $\Gamma_p\simeq 2.4$, shown by the thin black dashed line \cite{Neronov:2019ncc}. The model spectra are calculated using the AAFrag package \cite{Kachelriess:2019ifk,Koldobskiy:2021nld}. The normalization of the model $\gamma$-ray spectrum is adjusted to fit the $\gamma$-ray measurements in the energy range between \SI{10}{\giga\electronvolt} and \SI{3}{\tera\electronvolt}. The spectrum of neutrinos produced together with $\gamma$-rays is shown by the thin solid red line. One can see that the $\gamma$-ray and neutrino spectra are compatible. The hard multi-messenger spectrum of the Galactic Ridge may be due to the harder average cosmic ray spectrum in the inner galaxy \cite{Neronov:2015vua}, which may also originate from different properties of the cosmic ray source population or different energy dependence of the cosmic ray diffusion coefficient \cite{2015ApJ...815L..25G,Pagliaroli:2016lgg,Cataldo:2019qnz,Lipari:2018gzn}.

The simple pion decay model of the multi-messenger spectrum of the Galactic Ridge does not take into account the possible contribution of other emission components in the $\gamma$-ray spectrum. Apart from pion decay emission from cosmic ray interactions, the diffuse $\gamma$-ray flux from the Galactic Ridge may have a contribution from the inverse Compton scattering by cosmic ray electrons. In this case, the normalization of the pion decay $\gamma$-ray flux may be somewhat lower, compared to the pion-decay-only model. The resulting neutrino flux would also be lower. The locally observed cosmic ray spectrum has a "knee" feature at the energy $E \sim \SI{4}{\peta\electronvolt}$, possibly related to a high-energy cut-off in the galactic component of the cosmic ray spectrum \cite{KASCADEGrande:2011kpw,IceCube:2013ftu}. The ANTARES measurement is consistent with a possibility that also the cosmic ray spectrum in the Galactic Ridge has a cut-off at the knee energy, see Fig. \ref{fig:gamma_nu}. Overall, the limited statistics of the neutrino counterpart of the lower energy Fermi/LAT $\gamma$-ray signal from the Galactic Ridge is not yet sufficient for a reliable inference of the properties of the cosmic ray spectrum in the Galactic Ridge and for separating the leptonic and hadronic components of the $\gamma$-ray flux.

ANTARES measurements most tightly constrain the neutrino flux from the Galactic Ridge in the $\SIrange{10}{100}{\tera\electronvolt}$ energy band, where the $\gamma$-ray flux is poorly known. Nevertheless, within the pion decay model, the neutrino flux should have a $\gamma$-ray counterpart in the same energy range. The HESS telescope has previously reported the detection of a diffuse $\gamma$-ray signal from the inner Galactic Plane \cite{HESS:2014ree} at $E\gtrsim \SI{1}{\tera\electronvolt}$, but the spectral characteristics of the signal are not yet constrained and the properties of the signal above \SI{10}{\tera\electronvolt} are not clear. The Ridge region is located in the Southern sky and is largely inaccessible for existing wider field-of-view water-Cherenkov detector arrays such as HAWC and LHAASO, that are detecting $\gamma$-ray sources in the $\SIrange{10}{100}{\tera\electronvolt}$ range in the Northern hemisphere. Thus, it is currently not possible to find the $\gamma$-ray counterpart of the ANTARES excess in the direction of the Galactic Ridge.

Improved multi-messenger observations of the Galactic Ridge in neutrino + $\gamma$-ray channels will come with the next-generation detectors: KM3NeT \cite{KM3Net:2016zxf} which is under construction and has started taking data with a partial configuration and IceCube-Gen2 \cite{IceCube-Gen2:2020qha} for neutrinos, CTA \cite{CTAConsortium:2017dvg} and SWGO \cite{Abreu:2019ahw} for $\gamma$-rays. The southern site of the CTA observatory will be equipped with wider fields-of-view telescopes, compared to HESS, up to $\SI{9}{\degree}$ for the SST sub-array of CTA. This should facilitate the detection of diffuse emission from a $\sim \SI{4}{\degree}$ wide Galactic Ridge on top of the residual charged cosmic ray background in the telescope field of view. KM3NeT and SWGO will both have steradian-wide fields of view containing the entire Galactic Ridge source and sampling the complementary $\gamma$-ray and neutrino signals in the same energy range. Such a unique combination of multi-messenger data should enable the precise determination of the shapes of the $\gamma$-ray and neutrino spectra, the identification of hadronic and leptonic components of the multi-messenger flux, and the measurement of the spectra of cosmic ray protons/nuclei and electrons/positrons in the Galactic Ridge region.

\section*{Acknowledgement}

The authors acknowledge the financial support of the funding agencies:
Centre National de la Recherche Scientifique (CNRS), Commissariat \`a
l'\'ener\-gie atomique et aux \'energies alternatives (CEA),
Commission Europ\'eenne (FEDER fund and Marie Curie Program),
LabEx UnivEarthS (ANR-10-LABX-0023 and ANR-18-IDEX-0001),
R\'egion Alsace (contrat CPER), R\'egion Provence-Alpes-C\^ote d'Azur,
D\'e\-par\-tement du Var and Ville de La
Seyne-sur-Mer, France;
Bundesministerium f\"ur Bildung und Forschung
(BMBF), Germany; 
Istituto Nazionale di Fisica Nucleare (INFN), the European Union’s Horizon 2020 research and innovation programme under the Marie Sklodowska-Curie grant agreement No 754496, Italy;
Nederlandse organisatie voor Wetenschappelijk Onderzoek (NWO), the Netherlands;
Executive Unit for Financing Higher Education, Research, Development and Innovation (UEFISCDI), Romania;
Grants PID2021-124591NB-C41, -C42, -C43 funded by MCIN/AEI/ 10.13039/501100011033 and, as appropriate, by “ERDF A way of making Europe”, by the “European Union” or by the “European Union NextGenerationEU/PRTR”,  Programa de Planes Complementarios I+D+I (refs. ASFAE/2022/023, ASFAE/2022/014), Programa Prometeo (PROMETEO/2020/019) and GenT (refs. CIDEGENT/2018/034, /2019/043, /2020/049. /2021/23) of the Generalitat Valenciana, Junta de Andaluc\'{i}a (ref. P18-FR-5057), EU: MSC program (ref. 101025085), Programa Mar\'{i}a Zambrano (Spanish Ministry of Universities, funded by the European Union, NextGenerationEU), Spain;
Ministry of Higher Education, Scientific Research and Innovation, Morocco, and the Arab Fund for Economic and Social Development, Kuwait.
We also acknowledge the technical support of Ifremer, AIM and Foselev Marine
for the sea operation and the CC-IN2P3 for the computing facilities.

\bibliographystyle{elsarticle-num} 
\bibliography{references}

\end{document}

%% file: authors.tex
\author[IPHC,UHA]{A.~Albert}
\author[IFIC]{S.~Alves}
\author[UPC]{M.~Andr\'e}
\author[UPV]{M.~Ardid}
\author[UPV]{S.~Ardid}
\author[CPPM]{J.-J.~Aubert}
\author[APC]{J.~Aublin}
\author[APC]{B.~Baret}
\author[LAM]{S.~Basa}
\author[APC]{Y.~Becherini}
% ...

\address[IPHC]{Universit\'e de Strasbourg, CNRS,  IPHC UMR 7178, F-67000 Strasbourg, France}
\address[UHA]{Universit\'e de Haute Alsace, F-68100 Mulhouse, France}
\address[IFIC]{IFIC - Instituto de F\'isica Corpuscular (CSIC - Universitat de Val\`encia) c/ Catedr\'atico Jos\'e Beltr\'an, 2 E-46980 Paterna, Valencia, Spain}
\address[UPC]{Technical University of Catalonia, Laboratory of Applied Bioacoustics, Rambla Exposici\'o, 08800 Vilanova i la Geltr\'u, Barcelona, Spain}
\address[UPV]{Institut d'Investigaci\'o per a la Gesti\'o Integrada de les Zones Costaneres (IGIC) - Universitat Polit\`ecnica de Val\`encia. C/  Paranimf 1, 46730 Gandia, Spain}
\address[CPPM]{Aix Marseille Univ, CNRS/IN2P3, CPPM, Marseille, France}
\address[APC]{Universit\'e Paris Cit\'e, CNRS, Astroparticule et Cosmologie, F-75013 Paris, France}
\address[LAM]{Aix Marseille Univ, CNRS, CNES, LAM, Marseille, France }
% ...
